\documentclass[aps,prb,reprint,superscriptaddress,showpacs]{revtex4-1}
\pdfoutput=1
\usepackage{color}
\usepackage{graphicx}
\usepackage{amsmath}
\usepackage{amsfonts}
\usepackage{amssymb}
\usepackage{hyperref}
\allowdisplaybreaks
\usepackage{multirow}
\usepackage{braket}

\begin{document}

\author{Pieter~W.\ Claeys}
\email{PieterW.Claeys@UGent.be}
\affiliation{Institute for Theoretical Physics, University of Amsterdam, Science Park 904, 1098 XH Amsterdam, The Netherlands}
\affiliation{Ghent University, Department of Physics and Astronomy, Proeftuinstraat 86, 9000 Ghent, Belgium}
\affiliation{Ghent University, Center for Molecular Modeling, Technologiepark 903, 9052 Ghent, Belgium}

\author{Jean-S\'ebastien~Caux}
\affiliation{Institute for Theoretical Physics, University of Amsterdam, Science Park 904, 1098 XH Amsterdam, The Netherlands}

\title{Breaking the integrability of the Heisenberg model through periodic driving}

\begin{abstract}
We study the fate of interacting quantum systems which are periodically driven by switching back and forth between two integrable Hamiltonians. This provides an unconventional and tunable way of breaking integrability, in the sense that the stroboscopic time evolution will generally be described by a Floquet Hamiltonian which progressively becomes less integrable as the driving frequency is reduced. Here, we exemplify this idea in spin chains subjected to periodic switching between two integrable anisotropic Heisenberg Hamiltonians. We distinguish the integrability-breaking effects of resonant interactions and perturbative (local) interactions, and illustrate these by contrasting different measures of energy in Floquet states and through a study of level spacing statistics. This scenario is argued to be representative for general driven interacting integrable systems. 
\end{abstract}

\maketitle

\emph{Introduction.} -- Although the subject of driven quantum systems is quite an old one, the last couple of decades have witnessed an important increase of activity in the study of such systems, both in experimental and theoretical setups \cite{goldman_periodically_2014,bukov_universal_2015}. A general feature of interacting isolated systems subjected to periodic driving seems to be that they eventually heat up to an infinite temperature state, where the system loses all information about its initial state and all non-trivial correlations are lost \cite{dalessio_long-time_2014,lazarides_equilibrium_2014, ponte_periodically_2015,else_pre-thermal_2017}. This problem can be avoided by resorting to non-ergodic time evolution, where the existence of (approximate) conservation laws prevents an unlimited heating up and may lead to non-trivial steady states \cite{moessner_equilibration_2017}. Two well-studied classes where this is the case are integrable Floquet systems and Floquet systems exhibiting many-body localization \cite{dalessio_many-body_2013,lazarides_fate_2015,ponte_periodically_2015, khemani_phase_2016,bairey_driving-induced_2017}. In this work, we will focus on the former.

While no clear definition exists for quantum integrability in general \cite{caux_remarks_2011}, periodic driving complicates matters even more -- if the system is being driven by a time-dependent Hamiltonian which is integrable at each time, the resulting dynamics are governed by a Floquet Hamiltonian $\hat{H}_F$ which may or may not be integrable. Here a crucial distinction arises between systems which are integrable because they can be mapped to a non-interacting system, and truly interacting integrable systems. The former will always lead to a Floquet Hamiltonian which is similarly non-interacting and hence integrable, leading to non-ergodicity and a steady state which can be described by a periodic Generalized Gibbs Ensemble (PGGE) \cite{russomanno_periodic_2012,lazarides_periodic_2014,russomanno_entanglement_2016}, while the latter will only lead to integrable Floquet dynamics for extremely specific driving protocols \cite{prosen_time_1998,gritsev_integrable_2017}. Periodic driving using integrable Hamiltonians can hence be used to engineer non-integrable Floquet Hamiltonians. Remarkably, this breaking of integrability in periodically driven systems has not received much attention.

 Here, we investigate a driving protocol where we periodically switch between two Hamiltonians, both part of a one-parameter integrable family. The resulting Floquet Hamiltonian is integrable in the infinite-frequency limit, and two distinct interaction mechanisms are found to be responsible for a crossover from integrable (non-ergodic) to chaotic (ergodic) behaviour in finite systems. At high driving frequencies, the system does not have time to respond to changes in the time-dependent Hamiltonian, and will evolve as if governed by the time-averaged Hamiltonian, $\hat{H}_F=\hat{H}_{Avg}$ \cite{eckardt_high-frequency_2015,mikami_brillouin-wigner_2016,klarsfeld_baker-campbell-hausdorff_1989,blanes_magnus_2009,kuwahara_floquetmagnus_2016}, which can easily be chosen to be integrable. However, this no longer holds when moving away from this limit, and corrections on top of $\hat{H}_{Avg}$ need to be taken into account, breaking the integrability. Firstly, the non-commutativity of the driving Hamiltonians introduces perturbative local interations on top of the time-averaged Hamiltonian at finite frequencies. These can be captured by the Magnus expansion\cite{klarsfeld_baker-campbell-hausdorff_1989,blanes_magnus_2009,kuwahara_floquetmagnus_2016} and treated perturbatively, leading to a crossover at increasing perturbation strengths and driving period $T$. Secondly, periodically driven systems can exhibit resonant interactions\cite{hone_time-dependent_1997,eckardt_avoided-level-crossing_2008,hone_statistical_2009,dalessio_long-time_2014,goldman_periodically_2015,bukov_heating_2016,weinberg_adiabatic_nodate} -- states can interact by coupling to the driving, leading to strong interactions between states whose energies are separated by a multiple of the driving frequency $2\pi/T$. These cannot be described by local interactions, and are reflected in how the eigenvalues of the Floquet Hamiltonian are only defined up to integer shifts of $2\pi/T$.

A natural way of illustrating these effects is by comparing the quasienergies of Floquet states to their average energy per cycle. These coincide in the infinite-frequency limit and serve to highlight the deviation from the infinite-frequency Hamiltonian at finite frequencies, where both interaction mechanisms are shown to have a distinct effect. These are illustrated for a two-step driving protocol (or a periodic quench) by first connecting the derivatives of the Floquet quasienergies to expectation values of the time-averaged Hamiltonian, after which the influence of these interactions on the different contributions to the Floquet phases is made explicit.

\emph{Two-step driving protocol, the Magnus expansion and Floquet phases.} -- The Floquet theorem \cite{shirley_solution_1965,sambe_steady_1973,goldman_periodically_2014,bukov_universal_2015} allows the unitary evolution operator to be rewritten as
\begin{equation}
\hat{U}(t) = \hat{P}(t) e^{-i \hat{H}_F t},
\end{equation}
with $\hat{P}(t)$ a periodic unitary operator with the same period $T$ as the driving, $\hat{P}(t+T) = \hat{P}(t)$, and $\hat{H}_F$ the time-independent Floquet Hamiltonian. Furthermore, the fast-motion unitary operator $\hat{P}(t)$ reduces to the identity at stroboscopic times $t = n T, n \in \mathbb{N}$. The importance of this factorization is made clear when considering time-evolution over one cycle
\begin{equation}
\hat{U}_{F} \equiv \hat{U}(T) = e^{-i \hat{H}_F T},
\end{equation}
where at stroboscopic times the system behaves as if it evolves under the time-independent Floquet Hamiltonian. Simultaneously diagonalizing these operators then leads to
\begin{align}
\hat{H}_F = \sum_n \epsilon_n \ket{\phi_n}\bra{\phi_n}, \ \ \ 
\hat{U}_F = \sum_n e^{-i \theta_n} \ket{\phi_n}\bra{\phi_n},
\end{align}
where the eigenvalues of the Floquet Hamiltonian, also called quasienergies, are related to the Floquet phases as $\theta_n = \epsilon_n T$. The evolution within a single-period follows from the fast-motion operator, leading to states evolving as
\begin{equation}
\ket{\psi_n(t)} = e^{-i \epsilon_n t} \ket{\phi_n(t)}, \qquad \ket{\phi_n(t)} = \hat{P}(t) \ket{\phi_n},
\end{equation}
so that $\ket{\phi_n(t+T)} = \ket{\phi_n(t)}$. Plugging this in the time-dependent Schr\"odinger equation, the Floquet phases can be written as 
\begin{equation}
\theta_n = \int_{0}^T \braket{\phi_n(t) | \hat{H}(t) | \phi_n(t)} \mathrm{d}t -i\int_{0}^T \braket{\phi_n(t) | \partial_t | \phi_n(t)}  \mathrm{d}t,
\end{equation}
where the first term is the average energy of the state during a single cycle, leading to a dynamical phase contribution, while the second term describes a nonadiabatic (i.e. generalized) Berry phase \cite{grifoni_driven_1998}. Despite the apparent simplicity of these expressions, obtaining the Floquet Hamiltonian is a highly non-trivial task. Closed-form expressions are reserved for systems where the commutators of all involved Hamiltonians exhibit a clear structure \cite{gritsev_integrable_2017} (as e.g. in non-interacting systems \cite{russomanno_periodic_2012,lazarides_equilibrium_2014,russomanno_entanglement_2016,russomanno_spin_2017}), and numerical expressions are necessarily restricted to small system sizes due to the exponential scaling of the Hilbert space. These can be simplified by considering a two-step driving protocol, where within a single driving cycle we have
\begin{equation}
\hat{H}(t) = 
  \begin{cases}
 \hat{H}_1 &\text{for}\qquad 0 < t < \eta T, \\
  \hat{H}_2    & \text{for}\qquad \eta T < t < T,
  \end{cases}
\end{equation}
where $T$ is the total period of the driving and $\eta \in [0,1]$. This leads to a total evolution operator
\begin{equation}\label{floq:doublequench}
\hat{U}_F \equiv e^{-i \hat{H}_F T} = e^{-i (1-\eta) T \hat{H}_2} e^{-i \eta T \hat{H}_1},
\end{equation}
and as shown in the Supplementary Material, both contributions to the Floquet phases can then be simplified to single expectation values as
\begin{align}\label{floq:contr_phases}
\frac{1}{T}\int_{0}^T \braket{\phi_n(t) | \hat{H}(t) | \phi_n(t)} \mathrm{d}t &= \braket{\phi_n|\hat{H}_{Avg}|\phi_n}, \\
-\frac{i}{T}\int_{0}^T \braket{\phi_n(t) |  \partial_t | \phi_n(t)} \mathrm{d}t &= \braket{\phi_n|\hat{H}_F-\hat{H}_{Avg}|\phi_n},
\end{align}
with $\hat{H}_{Avg}=\eta \hat{H}_1 + (1-\eta) \hat{H}_2$. $\braket{\phi_n|\hat{H}_{Avg}|\phi_n}$ can be related to the energy absorbed during a single driving cycle\cite{dalessio_long-time_2014,rehn_how_2016}, and $\hat{H}_F-\hat{H}_{Avg}$ has a clear interpretation in the high-frequency limit. Here, the Magnus expansion\cite{klarsfeld_baker-campbell-hausdorff_1989,blanes_magnus_2009,kuwahara_floquetmagnus_2016} provides a series expansion of $\hat{H}_F$ in $T$, allowing the Floquet Hamiltonian to be approximated as $\hat{H}_F = \sum_{n=0}^{\infty}\hat{H}_F^{(n)}$, where the dominant term is precisely given by $\hat{H}_F^{(0)} = \hat{H}_{Avg} $ and the first higher-order corrections by
\begin{align}
\hat{H}_F^{(1)} = -\frac{iT}{4}[\hat{H}_{Avg},\hat{V}], \ \ \hat{H}_F^{(2)} =-\frac{T^2}{24}[[\hat{H}_{Avg},\hat{V}],\hat{V}],
\end{align}
with $\hat{V} = \eta \hat{H}_1-(1-\eta) \hat{H}_2$. Comparing with Eq. (\ref{floq:contr_phases}), it is clear that it is the higher-order terms in the Magnus expansion ($n \neq 0$) that give rise to the Berry phase. For two-step driving, these can also be connected to the Floquet phases as 
\begin{equation}
\frac{\theta_n}{T}=\bra{\phi_n} \hat{H}_F \ket{\phi_n}, \qquad \frac{\partial \theta_n}{\partial T} = \bra{\phi_n}\hat{H}_{avg} \ket{\phi_n},
\end{equation}
as similarly shown in the Supplementary Material, making use of techniques originated in Ref. \onlinecite{grifoni_driven_1998}.

\emph{Breaking integrability.} -- This distinction is particularly useful when investigating the deviation of the Floquet Hamiltonian from the time-averaged Hamiltonian. If the time-averaged Hamiltonian is chosen to be integrable, the Floquet Hamiltonian will be integrable in the infinite-frequency limit, but not at finite frequencies, leading to a crossover from integrable behaviour to non-integrable behaviour with increasing $T$. 

There are now two sources of interactions leading to the breaking of integrability. Firstly, the higher-order terms in the Magnus expansion introduce additional local interactions in the Floquet Hamiltonian, leading to a crossover with increasing perturbation strengths. Hence, once the higher-order terms (leading to non-negligible Berry phases) become relevant, the Floquet Hamiltonian will no longer behave as if it was integrable. Secondly, while the truncated Magnus expansion is known to provide a good approximation to the Floquet Hamiltonian \cite{mori_rigorous_2016,kuwahara_floquetmagnus_2016,abanin_effective_2017,abanin_rigorous_2017}, there can also be (highly nonlocal) resonant interactions in the system between energy levels separated by an integer times $2\pi/T$, which cannot be described by the local terms in the Magnus expansion \cite{hone_time-dependent_1997,eckardt_avoided-level-crossing_2008,hone_statistical_2009,dalessio_long-time_2014,goldman_periodically_2015,weinberg_adiabatic_nodate}.

This will be illustrated on the integrable anisotropic spin-$1/2$ Heisenberg chain \cite{bethe_zur_1931,orbach_linear_1958,korepin_quantum_1993} 
\begin{equation}
\hat{H}(t) = -J \sum_{i} \left[S_i^x S_{i+1}^x+S_i^y S_{i+1}^y + \Delta(t) S_i^z S_{i+1}^z\right].
\end{equation}
For numerical purposes, all calculations are restricted to periodic boundary conditions in the sector with total quasimomentum $k=0$, magnetization $m_z = 1/3$, parity $p=+1$, and $J=1$.
\begin{figure}
\includegraphics[width=\columnwidth]{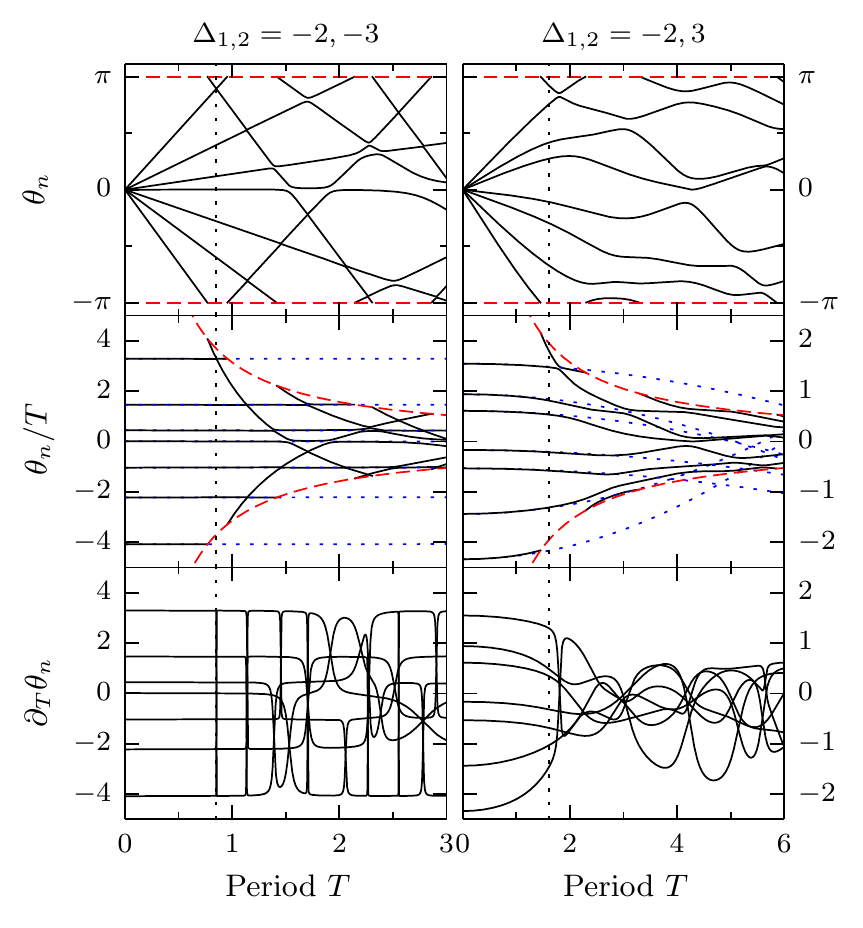}
\caption{Full spectrum of Floquet phases $\theta_n$, quasienergies $\theta_n/T$ and averaged energies $\partial_T \theta_n$ at different values of the driving period $T$ for $L=9$. The red lines mark $\pm \pi$ $(\theta_n)$ and $\pm \pi/T$ $(\theta_n/T)$, the blue lines are the PT2 results, and the vertical line marks $T=2 \pi/W$.  \label{fig:spec}}
\vspace{-\baselineskip}
\end{figure}
We first illustrate these effects in Fig. \ref{fig:spec} for small system sizes and driving between $\Delta_{1,2} = -2,-3$ and $\Delta_{1,2}=-2,3$. These values were chosen in order to best illustrate the different mechanisms at play, and are representative for a wider class of periodic quenches. 

In both cases, the phases $\theta_n$ are restricted to $[-\pi,\pi]$, the quasienergies $\theta_n/T$ to $[-\pi/T,\pi/T]$, and the averaged energies $\partial_T \theta_n$ are bounded by the extremal eigenvalues of $\hat{H}_{Avg}$. At small $T$, quasienergies and averaged energies equal the eigenvalues of the time-averaged Hamiltonian. For increasing $T$, two different behaviours can be noted, reflecting both sources of interactions. For $\Delta_{1,2} = -2,-3$, the energies remain approximately constant up until $T=2 \pi/W$, with $W$ the bandwidth of $\hat{H}_{Avg}$. At this point, the frequency of the driving equals the largest natural frequency in the system, and the extremal states can interact resonantly. In the spectrum of the Floquet Hamiltonian, this corresponds to one of these states crossing the edge of the Brillouin zone and undergoing an avoided crossing with the other state at $T= 2 \pi/W$. By further increasing $T$, more and more states will cross the edges of the Brillouin zone, leading to a multitude of avoided crossings in the phase and quasienergy spectrum, leading to the so-called folding of the Floquet spectrum. While the resolution does not always allow to visually distinguish between avoided and allowed level crossings, it can be observed from the abrupt transitions in $\partial_T \theta_n= \bra{\phi_n} \hat{H}_{Avg} \ket{\phi_n}$ that these are in fact avoided crossings, signifying interactions. Note that $\partial_T \theta_n$ remains approximately constant (modulo avoided crossings) for a range of $T > 2\pi/W$, indicating how most eigenstates of $\hat{H}_F$ remain well approximated by those of $\hat{H}_{Avg}$.

This can now be compared to the second driving protocol, $\Delta_{1,2} = -2,3$, where $\hat{V}$ is much larger. The aforementioned avoided crossings can similarly be observed for $T > 2 \pi/W$. However, both the quasienergies and the averaged energies deviate from their $T=0$ values well before these occur. This can be understood by applying second-order perturbation theory (PT2) on the Magnus expansion, leading to (see Supplementary Material)
\begin{equation}\label{floq:PT2}
\epsilon_n = \epsilon_n^{avg}-\frac{T^2}{96}\bra{\phi_n^{avg}}[[\hat{H}_{avg},\hat{V}],\hat{V}]\ket{\phi_n^{avg}}+\mathcal{O}(T^3).
\end{equation}
This approximation has also been presented in Fig. \ref{fig:spec}, where the corrections in the first driving protocol are negligible in the given range of $T$, whereas it is clear that the observed behaviour in the second protocol for $T < 2 \pi/W$ is well approximated by PT2 and can be attributed to the higher-order local terms arising in the Magnus expansion because of the non-commutativity of $\hat{H}_1$ and $\hat{H}_2$. 

\begin{figure}
\includegraphics{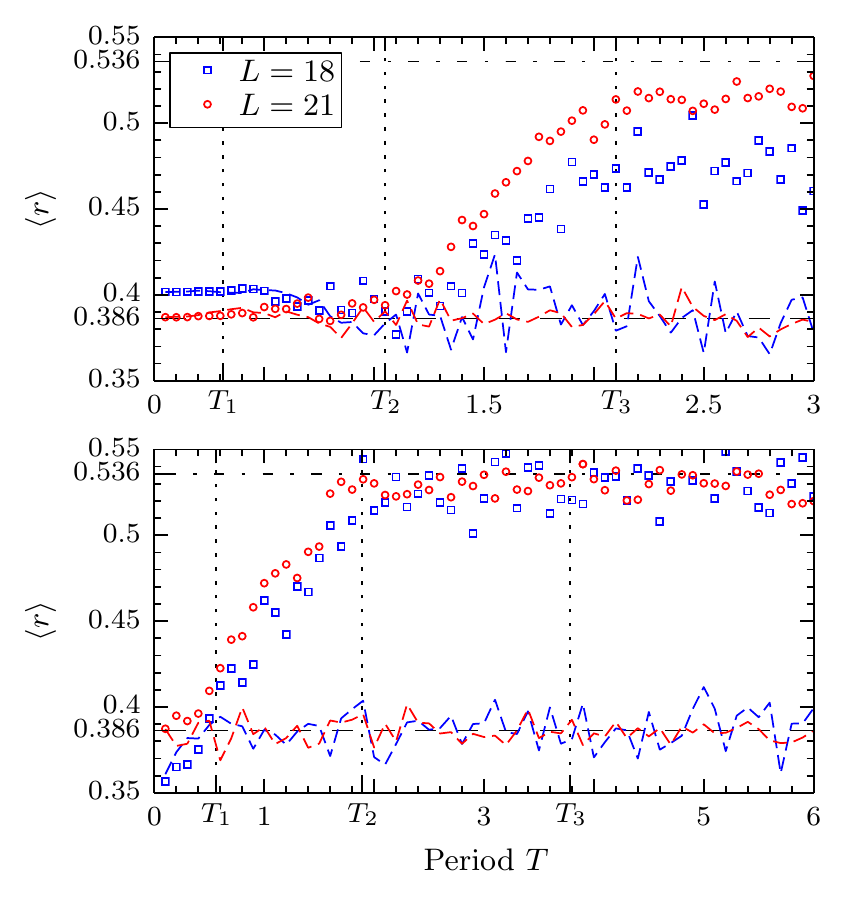}
\caption{Average values $\langle r(\theta_n/T) \rangle$ (symbols) and $\langle r(\partial_T \theta_n) \rangle$ (dashed lines) for periodic driving between two integrable Hamiltonians with $\Delta_{1,2} = -2, -3$ (Top) and $\Delta_{1,2} = -2, 3$ (Bottom).  \label{fig:r_int}}
\vspace{-\baselineskip}
\end{figure}

\emph{Level spacing statistics.} -- In order to quantify the effects of integrability-breaking for larger system sizes, it is customary to investigate the level statistics of the eigenvalue spectrum. Here, the Berry-Tabor conjecture can be used to distinguish the statistics for integrable and non-integrable Hamiltonians \cite{m._v._berry_level_1977,dalessio_quantum_2016}. Generally, it is expected that the level spacings of an integrable Hamiltonian behave according to Poissonian statistics (POI), and those of a non-integrable Hamiltonian satisfy the Wigner-Dyson statistics of a Gaussian orthogonal ensemble (GOE).  For a given set of ordered levels $\{E_n\}$, this can be quantified by defining $r$ as the ratio of two consecutive level spacings\cite{atas_distribution_2013},
\begin{equation}
r = \frac{\text{min}(s_n,s_{n+1})}{\text{max}(s_n,s_{n+1})} \in [0,1], \  s_n = E_{n+1}-E_{n},
\end{equation}
where GOE statistics would result in an average value of $\langle r \rangle_{GOE} \approx 0.535989$ and POI statistics in $\langle r \rangle_{POI} \approx 0.386295$. The underlying intuition is that non-integrable interactions lead to level repulsion and avoided crossings, characteristics of the GOE, as observed in Fig. \ref{fig:spec}.

\begin{figure}
\includegraphics[width=\columnwidth]{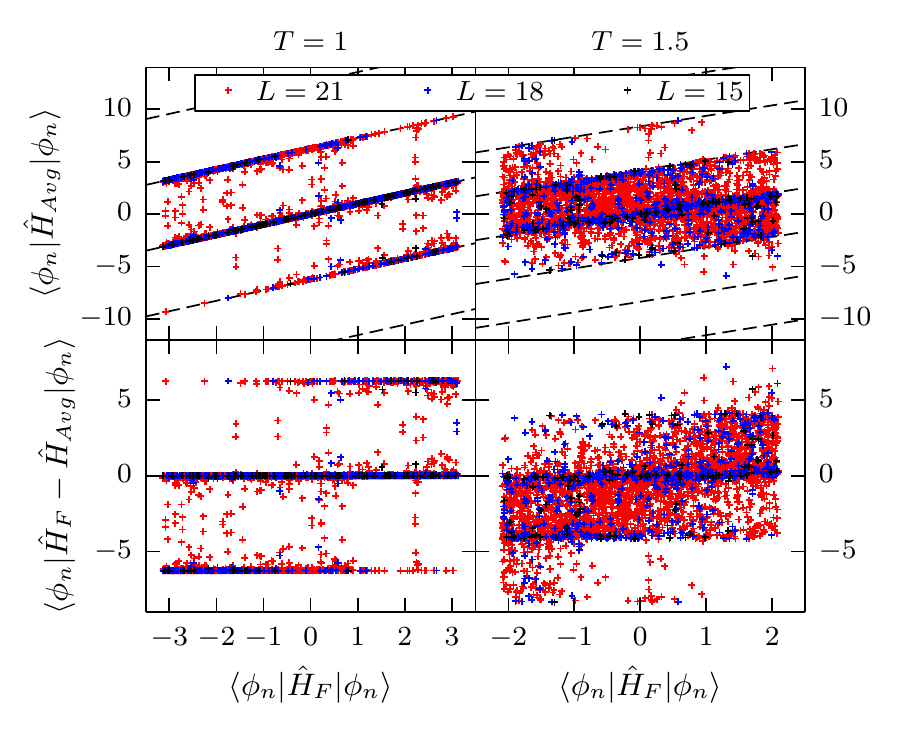}
\caption{Relation between the averaged energies $\langle \hat{H}_{Avg} \rangle = \partial_T \theta_n$ and quasienergies $\langle \hat{H}_{F} \rangle=\theta_n/T$ for driving $\Delta_{1,2}=-2,-3$ and $L=15,18,21$. $ \langle r(T=1) \rangle = 0.397$ and $\langle r(T=1.5) \rangle = 0.447$ for $L=21$.\label{fig:disp_top}}
\end{figure}

While it was already mentioned that the first resonant interactions occur around $T_1=\frac{2 \pi}{W}$, in Ref. \onlinecite{dalessio_long-time_2014} it was shown that the number of such interactions become statistically relevant between $T_2 = \frac{\pi}{\sigma}$ and $T_3 = \frac{2\pi}{\sigma}$, with $\sigma$ the variance of the spectrum of $\hat{H}_{Avg}$, which will be crucial in our analysis. In Fig. \ref{fig:r_int}, $\langle r \rangle$ is given for the two different driving protocols and different values of $T$. In order to better understand the effect of the crossings, this ratio is calculated both for the quasienergies and the averaged energies. 
In the top of Fig. \ref{fig:r_int}, at $T < T_1$ the Floquet Hamiltonian is well approximated by the time-averaged Hamiltonian, and both energies coincide and behave according to POI. At $T_1 < T < T_2$ both energies deviate, but no significant increase occurs. Then for $T_2 < T < T_3$ there is a clear crossover from POI to GOE due to resonant interactions, where $\langle r \rangle$ fluctuates around a fixed value for $T > T_3$, moving towards the GOE prediction with increasing system sizes. In Fig. \ref{fig:disp_top}, the effect on the contributions to the Floquet phases is illustrated. At small $T$, $\bra{\phi_n}\hat{H}_F \ket{\phi_n} = \bra{\phi_n}\hat{H}_{Avg} \ket{\phi_n} + \mathcal{O}(T^2)$, where crossing the edge of the Brillouin zone would result in integer shifts of $2\pi/T$. Resonant interactions then occur between states with similar quasienergies, leading to the mixing of values of $\bra{\phi_n}\hat{H}_{Avg} \ket{\phi_n}$ separated by such shifts, as can be observed in Fig. \ref{fig:disp_top}.
The effect of perturbative local interactions can be observed in the bottom of Fig. \ref{fig:r_int},  where an almost immediate increase in $\langle r \rangle$ occurs with increasing $T$, again moving from the POI to the GOE prediction, which is reached before resonant interactions become relevant.
\begin{figure}
\includegraphics[width=\columnwidth]{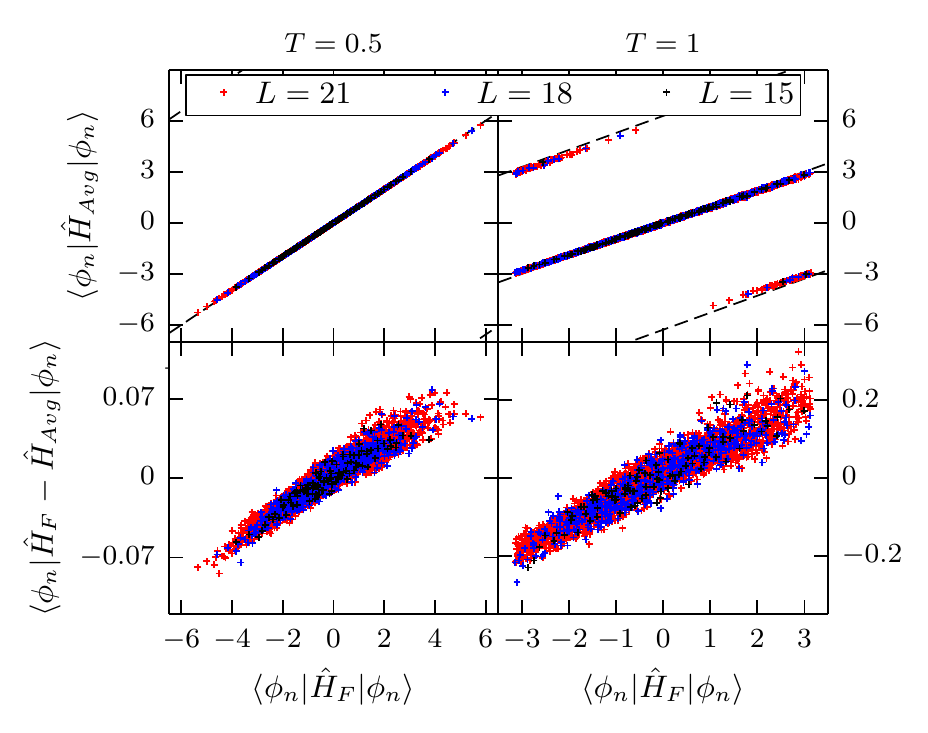}
\caption{Relation between the averaged energies $\langle \hat{H}_{Avg} \rangle = \partial_T \theta_n$ and quasienergies $\langle \hat{H}_{F} \rangle=\theta_n/T$ for driving $\Delta_{1,2}=-2,3$ and $L=15,18,21$. $ \langle r(T=0.5) \rangle = 0.409$ and $\langle r(T=1) \rangle = 0.472$ for $L=21$. \label{fig:disp_bottom}}
\end{figure}
In Fig. \ref{fig:disp_bottom}, the contributions to the Floquet phases are again made explicit. No significant amount of crossings between Brillouin zones due to resonant interactions occur, so the GOE statistics are entirely due to the perturbative local interactions $\hat{H}_F-\hat{H}_{Avg}$. Consistently with Eq. (\ref{floq:PT2}), the corrections (scaling as $T^2$) behave as the expectation values of local operators \cite{rigol_thermalization_2008,steinigeweg_eigenstate_2013,alba_eigenstate_2015,dalessio_quantum_2016}, exhibiting a relatively smooth dependence on the quasienergies.

Note that, while the effects in Fig. \ref{fig:disp_top} may seem more pronounced compared to Fig. \ref{fig:disp_bottom}, both represent systems where the level statistics indicate an equal measure of integrability-breaking.

\emph{Summary.} -- In this work, we have investigated the interactions responsible for the breaking of integrability in integrable interacting systems subjected to periodic driving. At high driving frequency, the Floquet Hamiltonian reduces to the time-averaged integrable Hamiltonian, where lowering the frequency introduces two kinds of interactions (resonant and perturbative local) responsible for the breaking of integrability. These were illustrated by contrasting two different measures of energy in Floquet states, which similarly highlight the deviation from the time-averaged Hamiltonian. While all calculations were restricted to the anisotropic Heisenberg model, the outlined reasoning does not depend on the specifics of the model at hand, and is expected to hold for a wider variety of interacting integrable systems including the Gaudin\cite{gaudin_bethe_2014} and Hubbard models\cite{essler_one-dimensional_2005}.

\emph{Acknowledgements.} -- We are grateful to S.E. Tapias Arze, E. Quinn, and V. Gritsev for valuable discussions and comments. P.W.C. acknowledges support from a Ph.D. fellowship and a travel grant for a long stay abroad at the University of Amsterdam from the Research Foundation Flanders (FWO Vlaanderen). J.-S. C. acknowledges support from the Foundation for Fundamental Research on Matter (FOM) and from the Netherlands Organization for Scientific Research (NWO). This work forms part of the activities of the Delta-Institute for Theoretical Physics (D-ITP).

\bibliography{FloquetBib.bib}

\begin{thebibliography}{47}%
\makeatletter
\providecommand \@ifxundefined [1]{%
 \@ifx{#1\undefined}
}%
\providecommand \@ifnum [1]{%
 \ifnum #1\expandafter \@firstoftwo
 \else \expandafter \@secondoftwo
 \fi
}%
\providecommand \@ifx [1]{%
 \ifx #1\expandafter \@firstoftwo
 \else \expandafter \@secondoftwo
 \fi
}%
\providecommand \natexlab [1]{#1}%
\providecommand \enquote  [1]{``#1''}%
\providecommand \bibnamefont  [1]{#1}%
\providecommand \bibfnamefont [1]{#1}%
\providecommand \citenamefont [1]{#1}%
\providecommand \href@noop [0]{\@secondoftwo}%
\providecommand \href [0]{\begingroup \@sanitize@url \@href}%
\providecommand \@href[1]{\@@startlink{#1}\@@href}%
\providecommand \@@href[1]{\endgroup#1\@@endlink}%
\providecommand \@sanitize@url [0]{\catcode `\\12\catcode `\$12\catcode
  `\&12\catcode `\#12\catcode `\^12\catcode `\_12\catcode `\%12\relax}%
\providecommand \@@startlink[1]{}%
\providecommand \@@endlink[0]{}%
\providecommand \url  [0]{\begingroup\@sanitize@url \@url }%
\providecommand \@url [1]{\endgroup\@href {#1}{\urlprefix }}%
\providecommand \urlprefix  [0]{URL }%
\providecommand \Eprint [0]{\href }%
\providecommand \doibase [0]{http://dx.doi.org/}%
\providecommand \selectlanguage [0]{\@gobble}%
\providecommand \bibinfo  [0]{\@secondoftwo}%
\providecommand \bibfield  [0]{\@secondoftwo}%
\providecommand \translation [1]{[#1]}%
\providecommand \BibitemOpen [0]{}%
\providecommand \bibitemStop [0]{}%
\providecommand \bibitemNoStop [0]{.\EOS\space}%
\providecommand \EOS [0]{\spacefactor3000\relax}%
\providecommand \BibitemShut  [1]{\csname bibitem#1\endcsname}%
\let\auto@bib@innerbib\@empty
\bibitem [{\citenamefont {Goldman}\ and\ \citenamefont
  {Dalibard}(2014)}]{goldman_periodically_2014}%
  \BibitemOpen
  \bibfield  {author} {\bibinfo {author} {\bibfnamefont {N.}~\bibnamefont
  {Goldman}}\ and\ \bibinfo {author} {\bibfnamefont {J.}~\bibnamefont
  {Dalibard}},\ }\href {\doibase 10.1103/PhysRevX.4.031027} {\bibfield
  {journal} {\bibinfo  {journal} {Phys. Rev. X}\ }\textbf {\bibinfo {volume}
  {4}},\ \bibinfo {pages} {031027} (\bibinfo {year} {2014})}\BibitemShut
  {NoStop}%
\bibitem [{\citenamefont {Bukov}\ \emph {et~al.}(2015)\citenamefont {Bukov},
  \citenamefont {D'Alessio},\ and\ \citenamefont
  {Polkovnikov}}]{bukov_universal_2015}%
  \BibitemOpen
  \bibfield  {author} {\bibinfo {author} {\bibfnamefont {M.}~\bibnamefont
  {Bukov}}, \bibinfo {author} {\bibfnamefont {L.}~\bibnamefont {D'Alessio}}, \
  and\ \bibinfo {author} {\bibfnamefont {A.}~\bibnamefont {Polkovnikov}},\
  }\href {\doibase 10.1080/00018732.2015.1055918} {\bibfield  {journal}
  {\bibinfo  {journal} {Adv. Phys.}\ }\textbf {\bibinfo {volume} {64}},\
  \bibinfo {pages} {139} (\bibinfo {year} {2015})}\BibitemShut {NoStop}%
\bibitem [{\citenamefont {D'Alessio}\ and\ \citenamefont
  {Rigol}(2014)}]{dalessio_long-time_2014}%
  \BibitemOpen
  \bibfield  {author} {\bibinfo {author} {\bibfnamefont {L.}~\bibnamefont
  {D'Alessio}}\ and\ \bibinfo {author} {\bibfnamefont {M.}~\bibnamefont
  {Rigol}},\ }\href {\doibase 10.1103/PhysRevX.4.041048} {\bibfield  {journal}
  {\bibinfo  {journal} {Phys. Rev. X}\ }\textbf {\bibinfo {volume} {4}},\
  \bibinfo {pages} {041048} (\bibinfo {year} {2014})}\BibitemShut {NoStop}%
\bibitem [{\citenamefont {Lazarides}\ \emph
  {et~al.}(2014{\natexlab{a}})\citenamefont {Lazarides}, \citenamefont {Das},\
  and\ \citenamefont {Moessner}}]{lazarides_equilibrium_2014}%
  \BibitemOpen
  \bibfield  {author} {\bibinfo {author} {\bibfnamefont {A.}~\bibnamefont
  {Lazarides}}, \bibinfo {author} {\bibfnamefont {A.}~\bibnamefont {Das}}, \
  and\ \bibinfo {author} {\bibfnamefont {R.}~\bibnamefont {Moessner}},\ }\href
  {\doibase 10.1103/PhysRevE.90.012110} {\bibfield  {journal} {\bibinfo
  {journal} {Phys. Rev. E}\ }\textbf {\bibinfo {volume} {90}},\ \bibinfo
  {pages} {012110} (\bibinfo {year} {2014}{\natexlab{a}})}\BibitemShut
  {NoStop}%
\bibitem [{\citenamefont {Ponte}\ \emph {et~al.}(2015)\citenamefont {Ponte},
  \citenamefont {Chandran}, \citenamefont {Papi{\'{c}}},\ and\ \citenamefont
  {Abanin}}]{ponte_periodically_2015}%
  \BibitemOpen
  \bibfield  {author} {\bibinfo {author} {\bibfnamefont {P.}~\bibnamefont
  {Ponte}}, \bibinfo {author} {\bibfnamefont {A.}~\bibnamefont {Chandran}},
  \bibinfo {author} {\bibfnamefont {Z.}~\bibnamefont {Papi{\'{c}}}}, \ and\
  \bibinfo {author} {\bibfnamefont {D.~A.}\ \bibnamefont {Abanin}},\ }\href
  {\doibase 10.1016/j.aop.2014.11.008} {\bibfield  {journal} {\bibinfo
  {journal} {Ann. Phys.}\ }\textbf {\bibinfo {volume} {353}},\ \bibinfo {pages}
  {196} (\bibinfo {year} {2015})}\BibitemShut {NoStop}%
\bibitem [{\citenamefont {Else}\ \emph {et~al.}(2017)\citenamefont {Else},
  \citenamefont {Bauer},\ and\ \citenamefont {Nayak}}]{else_pre-thermal_2017}%
  \BibitemOpen
  \bibfield  {author} {\bibinfo {author} {\bibfnamefont {D.~V.}\ \bibnamefont
  {Else}}, \bibinfo {author} {\bibfnamefont {B.}~\bibnamefont {Bauer}}, \ and\
  \bibinfo {author} {\bibfnamefont {C.}~\bibnamefont {Nayak}},\ }\href
  {\doibase 10.1103/PhysRevX.7.011026} {\bibfield  {journal} {\bibinfo
  {journal} {Phys. Rev. X}\ }\textbf {\bibinfo {volume} {7}},\ \bibinfo {pages}
  {011026} (\bibinfo {year} {2017})}\BibitemShut {NoStop}%
\bibitem [{\citenamefont {Moessner}\ and\ \citenamefont
  {Sondhi}(2017)}]{moessner_equilibration_2017}%
  \BibitemOpen
  \bibfield  {author} {\bibinfo {author} {\bibfnamefont {R.}~\bibnamefont
  {Moessner}}\ and\ \bibinfo {author} {\bibfnamefont {S.~L.}\ \bibnamefont
  {Sondhi}},\ }\href {\doibase 10.1038/nphys4106} {\bibfield  {journal}
  {\bibinfo  {journal} {Nat. Phys.}\ }\textbf {\bibinfo {volume} {13}},\
  \bibinfo {pages} {424} (\bibinfo {year} {2017})}\BibitemShut {NoStop}%
\bibitem [{\citenamefont {D'Alessio}\ and\ \citenamefont
  {Polkovnikov}(2013)}]{dalessio_many-body_2013}%
  \BibitemOpen
  \bibfield  {author} {\bibinfo {author} {\bibfnamefont {L.}~\bibnamefont
  {D'Alessio}}\ and\ \bibinfo {author} {\bibfnamefont {A.}~\bibnamefont
  {Polkovnikov}},\ }\href {\doibase 10.1016/j.aop.2013.02.011} {\bibfield
  {journal} {\bibinfo  {journal} {Ann. Phys.}\ }\textbf {\bibinfo {volume}
  {333}},\ \bibinfo {pages} {19} (\bibinfo {year} {2013})}\BibitemShut
  {NoStop}%
\bibitem [{\citenamefont {Lazarides}\ \emph {et~al.}(2015)\citenamefont
  {Lazarides}, \citenamefont {Das},\ and\ \citenamefont
  {Moessner}}]{lazarides_fate_2015}%
  \BibitemOpen
  \bibfield  {author} {\bibinfo {author} {\bibfnamefont {A.}~\bibnamefont
  {Lazarides}}, \bibinfo {author} {\bibfnamefont {A.}~\bibnamefont {Das}}, \
  and\ \bibinfo {author} {\bibfnamefont {R.}~\bibnamefont {Moessner}},\ }\href
  {\doibase 10.1103/PhysRevLett.115.030402} {\bibfield  {journal} {\bibinfo
  {journal} {Phys. Rev. Lett.}\ }\textbf {\bibinfo {volume} {115}},\ \bibinfo
  {pages} {030402} (\bibinfo {year} {2015})}\BibitemShut {NoStop}%
\bibitem [{\citenamefont {Khemani}\ \emph {et~al.}(2016)\citenamefont
  {Khemani}, \citenamefont {Lazarides}, \citenamefont {Moessner},\ and\
  \citenamefont {Sondhi}}]{khemani_phase_2016}%
  \BibitemOpen
  \bibfield  {author} {\bibinfo {author} {\bibfnamefont {V.}~\bibnamefont
  {Khemani}}, \bibinfo {author} {\bibfnamefont {A.}~\bibnamefont {Lazarides}},
  \bibinfo {author} {\bibfnamefont {R.}~\bibnamefont {Moessner}}, \ and\
  \bibinfo {author} {\bibfnamefont {S.}~\bibnamefont {Sondhi}},\ }\href
  {\doibase 10.1103/PhysRevLett.116.250401} {\bibfield  {journal} {\bibinfo
  {journal} {Phys. Rev. Lett.}\ }\textbf {\bibinfo {volume} {116}},\ \bibinfo
  {pages} {250401} (\bibinfo {year} {2016})}\BibitemShut {NoStop}%
\bibitem [{\citenamefont {Bairey}\ \emph {et~al.}(2017)\citenamefont {Bairey},
  \citenamefont {Refael},\ and\ \citenamefont
  {Lindner}}]{bairey_driving-induced_2017}%
  \BibitemOpen
  \bibfield  {author} {\bibinfo {author} {\bibfnamefont {E.}~\bibnamefont
  {Bairey}}, \bibinfo {author} {\bibfnamefont {G.}~\bibnamefont {Refael}}, \
  and\ \bibinfo {author} {\bibfnamefont {N.~H.}\ \bibnamefont {Lindner}},\
  }\href {\doibase 10.1103/PhysRevB.96.020201} {\bibfield  {journal} {\bibinfo
  {journal} {Phys. Rev. B}\ }\textbf {\bibinfo {volume} {96}},\ \bibinfo
  {pages} {020201} (\bibinfo {year} {2017})}\BibitemShut {NoStop}%
\bibitem [{\citenamefont {Caux}\ and\ \citenamefont
  {Mossel}(2011)}]{caux_remarks_2011}%
  \BibitemOpen
  \bibfield  {author} {\bibinfo {author} {\bibfnamefont {J.-S.}\ \bibnamefont
  {Caux}}\ and\ \bibinfo {author} {\bibfnamefont {J.}~\bibnamefont {Mossel}},\
  }\href {\doibase 10.1088/1742-5468/2011/02/P02023} {\bibfield  {journal}
  {\bibinfo  {journal} {J. Stat. Mech: Th. Exp.}\ }\textbf {\bibinfo {volume}
  {2011}},\ \bibinfo {pages} {P02023} (\bibinfo {year} {2011})}\BibitemShut
  {NoStop}%
\bibitem [{\citenamefont {Russomanno}\ \emph {et~al.}(2012)\citenamefont
  {Russomanno}, \citenamefont {Silva},\ and\ \citenamefont
  {Santoro}}]{russomanno_periodic_2012}%
  \BibitemOpen
  \bibfield  {author} {\bibinfo {author} {\bibfnamefont {A.}~\bibnamefont
  {Russomanno}}, \bibinfo {author} {\bibfnamefont {A.}~\bibnamefont {Silva}}, \
  and\ \bibinfo {author} {\bibfnamefont {G.~E.}\ \bibnamefont {Santoro}},\
  }\href {\doibase 10.1103/PhysRevLett.109.257201} {\bibfield  {journal}
  {\bibinfo  {journal} {Phys. Rev. Lett.}\ }\textbf {\bibinfo {volume} {109}},\
  \bibinfo {pages} {257201} (\bibinfo {year} {2012})}\BibitemShut {NoStop}%
\bibitem [{\citenamefont {Lazarides}\ \emph
  {et~al.}(2014{\natexlab{b}})\citenamefont {Lazarides}, \citenamefont {Das},\
  and\ \citenamefont {Moessner}}]{lazarides_periodic_2014}%
  \BibitemOpen
  \bibfield  {author} {\bibinfo {author} {\bibfnamefont {A.}~\bibnamefont
  {Lazarides}}, \bibinfo {author} {\bibfnamefont {A.}~\bibnamefont {Das}}, \
  and\ \bibinfo {author} {\bibfnamefont {R.}~\bibnamefont {Moessner}},\ }\href
  {\doibase 10.1103/PhysRevLett.112.150401} {\bibfield  {journal} {\bibinfo
  {journal} {Phys. Rev. Lett.}\ }\textbf {\bibinfo {volume} {112}},\ \bibinfo
  {pages} {150401} (\bibinfo {year} {2014}{\natexlab{b}})}\BibitemShut
  {NoStop}%
\bibitem [{\citenamefont {Russomanno}\ \emph {et~al.}(2016)\citenamefont
  {Russomanno}, \citenamefont {Santoro},\ and\ \citenamefont
  {Fazio}}]{russomanno_entanglement_2016}%
  \BibitemOpen
  \bibfield  {author} {\bibinfo {author} {\bibfnamefont {A.}~\bibnamefont
  {Russomanno}}, \bibinfo {author} {\bibfnamefont {G.~E.}\ \bibnamefont
  {Santoro}}, \ and\ \bibinfo {author} {\bibfnamefont {R.}~\bibnamefont
  {Fazio}},\ }\href {\doibase 10.1088/1742-5468/2016/07/073101} {\bibfield
  {journal} {\bibinfo  {journal} {J. Stat. Mech: Th. Exp.}\ }\textbf {\bibinfo
  {volume} {2016}},\ \bibinfo {pages} {073101} (\bibinfo {year}
  {2016})}\BibitemShut {NoStop}%
\bibitem [{\citenamefont {Prosen}(1998)}]{prosen_time_1998}%
  \BibitemOpen
  \bibfield  {author} {\bibinfo {author} {\bibfnamefont {T.}~\bibnamefont
  {Prosen}},\ }\href {\doibase 10.1103/PhysRevLett.80.1808} {\bibfield
  {journal} {\bibinfo  {journal} {Phys. Rev. Lett.}\ }\textbf {\bibinfo
  {volume} {80}},\ \bibinfo {pages} {1808} (\bibinfo {year}
  {1998})}\BibitemShut {NoStop}%
\bibitem [{\citenamefont {Gritsev}\ and\ \citenamefont
  {Polkovnikov}(2017)}]{gritsev_integrable_2017}%
  \BibitemOpen
  \bibfield  {author} {\bibinfo {author} {\bibfnamefont {V.}~\bibnamefont
  {Gritsev}}\ and\ \bibinfo {author} {\bibfnamefont {A.}~\bibnamefont
  {Polkovnikov}},\ }\href {\doibase 10.21468/SciPostPhys.2.3.021} {\bibfield
  {journal} {\bibinfo  {journal} {SciPost Phys.}\ }\textbf {\bibinfo {volume}
  {2}},\ \bibinfo {pages} {021} (\bibinfo {year} {2017})}\BibitemShut {NoStop}%
\bibitem [{\citenamefont {Eckardt}\ and\ \citenamefont
  {Anisimovas}(2015)}]{eckardt_high-frequency_2015}%
  \BibitemOpen
  \bibfield  {author} {\bibinfo {author} {\bibfnamefont {A.}~\bibnamefont
  {Eckardt}}\ and\ \bibinfo {author} {\bibfnamefont {E.}~\bibnamefont
  {Anisimovas}},\ }\href {\doibase 10.1088/1367-2630/17/9/093039} {\bibfield
  {journal} {\bibinfo  {journal} {New J. Phys.}\ }\textbf {\bibinfo {volume}
  {17}},\ \bibinfo {pages} {093039} (\bibinfo {year} {2015})}\BibitemShut
  {NoStop}%
\bibitem [{\citenamefont {Mikami}\ \emph {et~al.}(2016)\citenamefont {Mikami},
  \citenamefont {Kitamura}, \citenamefont {Yasuda}, \citenamefont {Tsuji},
  \citenamefont {Oka},\ and\ \citenamefont
  {Aoki}}]{mikami_brillouin-wigner_2016}%
  \BibitemOpen
  \bibfield  {author} {\bibinfo {author} {\bibfnamefont {T.}~\bibnamefont
  {Mikami}}, \bibinfo {author} {\bibfnamefont {S.}~\bibnamefont {Kitamura}},
  \bibinfo {author} {\bibfnamefont {K.}~\bibnamefont {Yasuda}}, \bibinfo
  {author} {\bibfnamefont {N.}~\bibnamefont {Tsuji}}, \bibinfo {author}
  {\bibfnamefont {T.}~\bibnamefont {Oka}}, \ and\ \bibinfo {author}
  {\bibfnamefont {H.}~\bibnamefont {Aoki}},\ }\href {\doibase
  10.1103/PhysRevB.93.144307} {\bibfield  {journal} {\bibinfo  {journal} {Phys.
  Rev. B}\ }\textbf {\bibinfo {volume} {93}},\ \bibinfo {pages} {144307}
  (\bibinfo {year} {2016})}\BibitemShut {NoStop}%
\bibitem [{\citenamefont {Klarsfeld}\ and\ \citenamefont
  {Oteo}(1989)}]{klarsfeld_baker-campbell-hausdorff_1989}%
  \BibitemOpen
  \bibfield  {author} {\bibinfo {author} {\bibfnamefont {S.}~\bibnamefont
  {Klarsfeld}}\ and\ \bibinfo {author} {\bibfnamefont {J.~A.}\ \bibnamefont
  {Oteo}},\ }\href {\doibase 10.1088/0305-4470/22/21/018} {\bibfield  {journal}
  {\bibinfo  {journal} {J. Phys. A: Math. Gen.}\ }\textbf {\bibinfo {volume}
  {22}},\ \bibinfo {pages} {4565} (\bibinfo {year} {1989})}\BibitemShut
  {NoStop}%
\bibitem [{\citenamefont {Blanes}\ \emph {et~al.}(2009)\citenamefont {Blanes},
  \citenamefont {Casas}, \citenamefont {Oteo},\ and\ \citenamefont
  {Ros}}]{blanes_magnus_2009}%
  \BibitemOpen
  \bibfield  {author} {\bibinfo {author} {\bibfnamefont {S.}~\bibnamefont
  {Blanes}}, \bibinfo {author} {\bibfnamefont {F.}~\bibnamefont {Casas}},
  \bibinfo {author} {\bibfnamefont {J.~A.}\ \bibnamefont {Oteo}}, \ and\
  \bibinfo {author} {\bibfnamefont {J.}~\bibnamefont {Ros}},\ }\href {\doibase
  10.1016/j.physrep.2008.11.001} {\bibfield  {journal} {\bibinfo  {journal}
  {Phys. Rep.}\ }\textbf {\bibinfo {volume} {470}},\ \bibinfo {pages} {151}
  (\bibinfo {year} {2009})}\BibitemShut {NoStop}%
\bibitem [{\citenamefont {Kuwahara}\ \emph {et~al.}(2016)\citenamefont
  {Kuwahara}, \citenamefont {Mori},\ and\ \citenamefont
  {Saito}}]{kuwahara_floquetmagnus_2016}%
  \BibitemOpen
  \bibfield  {author} {\bibinfo {author} {\bibfnamefont {T.}~\bibnamefont
  {Kuwahara}}, \bibinfo {author} {\bibfnamefont {T.}~\bibnamefont {Mori}}, \
  and\ \bibinfo {author} {\bibfnamefont {K.}~\bibnamefont {Saito}},\ }\href
  {\doibase 10.1016/j.aop.2016.01.012} {\bibfield  {journal} {\bibinfo
  {journal} {Ann. Phys.}\ }\textbf {\bibinfo {volume} {367}},\ \bibinfo {pages}
  {96} (\bibinfo {year} {2016})}\BibitemShut {NoStop}%
\bibitem [{\citenamefont {Hone}\ \emph {et~al.}(1997)\citenamefont {Hone},
  \citenamefont {Ketzmerick},\ and\ \citenamefont
  {Kohn}}]{hone_time-dependent_1997}%
  \BibitemOpen
  \bibfield  {author} {\bibinfo {author} {\bibfnamefont {D.~W.}\ \bibnamefont
  {Hone}}, \bibinfo {author} {\bibfnamefont {R.}~\bibnamefont {Ketzmerick}}, \
  and\ \bibinfo {author} {\bibfnamefont {W.}~\bibnamefont {Kohn}},\ }\href
  {\doibase 10.1103/PhysRevA.56.4045} {\bibfield  {journal} {\bibinfo
  {journal} {Phys. Rev. A}\ }\textbf {\bibinfo {volume} {56}},\ \bibinfo
  {pages} {4045} (\bibinfo {year} {1997})}\BibitemShut {NoStop}%
\bibitem [{\citenamefont {Eckardt}\ and\ \citenamefont
  {Holthaus}(2008)}]{eckardt_avoided-level-crossing_2008}%
  \BibitemOpen
  \bibfield  {author} {\bibinfo {author} {\bibfnamefont {A.}~\bibnamefont
  {Eckardt}}\ and\ \bibinfo {author} {\bibfnamefont {M.}~\bibnamefont
  {Holthaus}},\ }\href {\doibase 10.1103/PhysRevLett.101.245302} {\bibfield
  {journal} {\bibinfo  {journal} {Phys. Rev. Lett.}\ }\textbf {\bibinfo
  {volume} {101}},\ \bibinfo {pages} {245302} (\bibinfo {year}
  {2008})}\BibitemShut {NoStop}%
\bibitem [{\citenamefont {Hone}\ \emph {et~al.}(2009)\citenamefont {Hone},
  \citenamefont {Ketzmerick},\ and\ \citenamefont
  {Kohn}}]{hone_statistical_2009}%
  \BibitemOpen
  \bibfield  {author} {\bibinfo {author} {\bibfnamefont {D.~W.}\ \bibnamefont
  {Hone}}, \bibinfo {author} {\bibfnamefont {R.}~\bibnamefont {Ketzmerick}}, \
  and\ \bibinfo {author} {\bibfnamefont {W.}~\bibnamefont {Kohn}},\ }\href
  {\doibase 10.1103/PhysRevE.79.051129} {\bibfield  {journal} {\bibinfo
  {journal} {Phys. Rev. E}\ }\textbf {\bibinfo {volume} {79}},\ \bibinfo
  {pages} {051129} (\bibinfo {year} {2009})}\BibitemShut {NoStop}%
\bibitem [{\citenamefont {Goldman}\ \emph {et~al.}(2015)\citenamefont
  {Goldman}, \citenamefont {Dalibard}, \citenamefont {Aidelsburger},\ and\
  \citenamefont {Cooper}}]{goldman_periodically_2015}%
  \BibitemOpen
  \bibfield  {author} {\bibinfo {author} {\bibfnamefont {N.}~\bibnamefont
  {Goldman}}, \bibinfo {author} {\bibfnamefont {J.}~\bibnamefont {Dalibard}},
  \bibinfo {author} {\bibfnamefont {M.}~\bibnamefont {Aidelsburger}}, \ and\
  \bibinfo {author} {\bibfnamefont {N.~R.}\ \bibnamefont {Cooper}},\ }\href
  {\doibase 10.1103/PhysRevA.91.033632} {\bibfield  {journal} {\bibinfo
  {journal} {Phys. Rev. A}\ }\textbf {\bibinfo {volume} {91}},\ \bibinfo
  {pages} {033632} (\bibinfo {year} {2015})}\BibitemShut {NoStop}%
\bibitem [{\citenamefont {Bukov}\ \emph {et~al.}(2016)\citenamefont {Bukov},
  \citenamefont {Heyl}, \citenamefont {Huse},\ and\ \citenamefont
  {Polkovnikov}}]{bukov_heating_2016}%
  \BibitemOpen
  \bibfield  {author} {\bibinfo {author} {\bibfnamefont {M.}~\bibnamefont
  {Bukov}}, \bibinfo {author} {\bibfnamefont {M.}~\bibnamefont {Heyl}},
  \bibinfo {author} {\bibfnamefont {D.~A.}\ \bibnamefont {Huse}}, \ and\
  \bibinfo {author} {\bibfnamefont {A.}~\bibnamefont {Polkovnikov}},\ }\href
  {\doibase 10.1103/PhysRevB.93.155132} {\bibfield  {journal} {\bibinfo
  {journal} {Phys. Rev. B}\ }\textbf {\bibinfo {volume} {93}},\ \bibinfo
  {pages} {155132} (\bibinfo {year} {2016})}\BibitemShut {NoStop}%
\bibitem [{\citenamefont {Weinberg}\ \emph {et~al.}(2017)\citenamefont
  {Weinberg}, \citenamefont {Bukov}, \citenamefont {D{\textquoteright}Alessio},
  \citenamefont {Polkovnikov}, \citenamefont {Vajna},\ and\ \citenamefont
  {Kolodrubetz}}]{weinberg_adiabatic_nodate}%
  \BibitemOpen
  \bibfield  {author} {\bibinfo {author} {\bibfnamefont {P.}~\bibnamefont
  {Weinberg}}, \bibinfo {author} {\bibfnamefont {M.}~\bibnamefont {Bukov}},
  \bibinfo {author} {\bibfnamefont {L.}~\bibnamefont
  {D{\textquoteright}Alessio}}, \bibinfo {author} {\bibfnamefont
  {A.}~\bibnamefont {Polkovnikov}}, \bibinfo {author} {\bibfnamefont
  {S.}~\bibnamefont {Vajna}}, \ and\ \bibinfo {author} {\bibfnamefont
  {M.}~\bibnamefont {Kolodrubetz}},\ }\href {\doibase
  10.1016/j.physrep.2017.05.003} {\bibfield  {journal} {\bibinfo  {journal}
  {Phys. Rep.}\ }\textbf {\bibinfo {volume} {688}},\ \bibinfo {pages} {1 }
  (\bibinfo {year} {2017})}\BibitemShut {NoStop}%
\bibitem [{\citenamefont {Shirley}(1965)}]{shirley_solution_1965}%
  \BibitemOpen
  \bibfield  {author} {\bibinfo {author} {\bibfnamefont {J.~H.}\ \bibnamefont
  {Shirley}},\ }\href {\doibase 10.1103/PhysRev.138.B979} {\bibfield  {journal}
  {\bibinfo  {journal} {Phys. Rev.}\ }\textbf {\bibinfo {volume} {138}},\
  \bibinfo {pages} {B979} (\bibinfo {year} {1965})}\BibitemShut {NoStop}%
\bibitem [{\citenamefont {Sambe}(1973)}]{sambe_steady_1973}%
  \BibitemOpen
  \bibfield  {author} {\bibinfo {author} {\bibfnamefont {H.}~\bibnamefont
  {Sambe}},\ }\href {\doibase 10.1103/PhysRevA.7.2203} {\bibfield  {journal}
  {\bibinfo  {journal} {Phys. Rev. A}\ }\textbf {\bibinfo {volume} {7}},\
  \bibinfo {pages} {2203} (\bibinfo {year} {1973})}\BibitemShut {NoStop}%
\bibitem [{\citenamefont {Grifoni}\ and\ \citenamefont
  {H{\"{a}}nggi}(1998)}]{grifoni_driven_1998}%
  \BibitemOpen
  \bibfield  {author} {\bibinfo {author} {\bibfnamefont {M.}~\bibnamefont
  {Grifoni}}\ and\ \bibinfo {author} {\bibfnamefont {P.}~\bibnamefont
  {H{\"{a}}nggi}},\ }\href {\doibase 10.1016/S0370-1573(98)00022-2} {\bibfield
  {journal} {\bibinfo  {journal} {Phys. Rep.}\ }\textbf {\bibinfo {volume}
  {304}},\ \bibinfo {pages} {229} (\bibinfo {year} {1998})}\BibitemShut
  {NoStop}%
\bibitem [{\citenamefont {Russomanno}\ \emph {et~al.}(2017)\citenamefont
  {Russomanno}, \citenamefont {Friedman},\ and\ \citenamefont
  {Dalla~Torre}}]{russomanno_spin_2017}%
  \BibitemOpen
  \bibfield  {author} {\bibinfo {author} {\bibfnamefont {A.}~\bibnamefont
  {Russomanno}}, \bibinfo {author} {\bibfnamefont {B.-e.}\ \bibnamefont
  {Friedman}}, \ and\ \bibinfo {author} {\bibfnamefont {E.~G.}\ \bibnamefont
  {Dalla~Torre}},\ }\href {\doibase 10.1103/PhysRevB.96.045422} {\bibfield
  {journal} {\bibinfo  {journal} {Phys. Rev. B}\ }\textbf {\bibinfo {volume}
  {96}},\ \bibinfo {pages} {045422} (\bibinfo {year} {2017})}\BibitemShut
  {NoStop}%
\bibitem [{\citenamefont {Rehn}\ \emph {et~al.}(2016)\citenamefont {Rehn},
  \citenamefont {Lazarides}, \citenamefont {Pollmann},\ and\ \citenamefont
  {Moessner}}]{rehn_how_2016}%
  \BibitemOpen
  \bibfield  {author} {\bibinfo {author} {\bibfnamefont {J.}~\bibnamefont
  {Rehn}}, \bibinfo {author} {\bibfnamefont {A.}~\bibnamefont {Lazarides}},
  \bibinfo {author} {\bibfnamefont {F.}~\bibnamefont {Pollmann}}, \ and\
  \bibinfo {author} {\bibfnamefont {R.}~\bibnamefont {Moessner}},\ }\href
  {\doibase 10.1103/PhysRevB.94.020201} {\bibfield  {journal} {\bibinfo
  {journal} {Phys. Rev. B}\ }\textbf {\bibinfo {volume} {94}},\ \bibinfo
  {pages} {020201} (\bibinfo {year} {2016})}\BibitemShut {NoStop}%
\bibitem [{\citenamefont {Mori}\ \emph {et~al.}(2016)\citenamefont {Mori},
  \citenamefont {Kuwahara},\ and\ \citenamefont {Saito}}]{mori_rigorous_2016}%
  \BibitemOpen
  \bibfield  {author} {\bibinfo {author} {\bibfnamefont {T.}~\bibnamefont
  {Mori}}, \bibinfo {author} {\bibfnamefont {T.}~\bibnamefont {Kuwahara}}, \
  and\ \bibinfo {author} {\bibfnamefont {K.}~\bibnamefont {Saito}},\ }\href
  {\doibase 10.1103/PhysRevLett.116.120401} {\bibfield  {journal} {\bibinfo
  {journal} {Phys. Rev. Lett.}\ }\textbf {\bibinfo {volume} {116}},\ \bibinfo
  {pages} {120401} (\bibinfo {year} {2016})}\BibitemShut {NoStop}%
\bibitem [{\citenamefont {Abanin}\ \emph
  {et~al.}(2017{\natexlab{a}})\citenamefont {Abanin}, \citenamefont {De~Roeck},
  \citenamefont {Ho},\ and\ \citenamefont {Huveneers}}]{abanin_effective_2017}%
  \BibitemOpen
  \bibfield  {author} {\bibinfo {author} {\bibfnamefont {D.~A.}\ \bibnamefont
  {Abanin}}, \bibinfo {author} {\bibfnamefont {W.}~\bibnamefont {De~Roeck}},
  \bibinfo {author} {\bibfnamefont {W.~W.}\ \bibnamefont {Ho}}, \ and\ \bibinfo
  {author} {\bibfnamefont {F.}~\bibnamefont {Huveneers}},\ }\href {\doibase
  10.1103/PhysRevB.95.014112} {\bibfield  {journal} {\bibinfo  {journal} {Phys.
  Rev. B}\ }\textbf {\bibinfo {volume} {95}},\ \bibinfo {pages} {014112}
  (\bibinfo {year} {2017}{\natexlab{a}})}\BibitemShut {NoStop}%
\bibitem [{\citenamefont {Abanin}\ \emph
  {et~al.}(2017{\natexlab{b}})\citenamefont {Abanin}, \citenamefont {De~Roeck},
  \citenamefont {Ho},\ and\ \citenamefont {Huveneers}}]{abanin_rigorous_2017}%
  \BibitemOpen
  \bibfield  {author} {\bibinfo {author} {\bibfnamefont {D.}~\bibnamefont
  {Abanin}}, \bibinfo {author} {\bibfnamefont {W.}~\bibnamefont {De~Roeck}},
  \bibinfo {author} {\bibfnamefont {W.~W.}\ \bibnamefont {Ho}}, \ and\ \bibinfo
  {author} {\bibfnamefont {F.}~\bibnamefont {Huveneers}},\ }\href {\doibase
  10.1007/s00220-017-2930-x} {\bibfield  {journal} {\bibinfo  {journal}
  {Commun. Math. Phys.}\ }\textbf {\bibinfo {volume} {354}},\ \bibinfo {pages}
  {809} (\bibinfo {year} {2017}{\natexlab{b}})}\BibitemShut {NoStop}%
\bibitem [{\citenamefont {Bethe}(1931)}]{bethe_zur_1931}%
  \BibitemOpen
  \bibfield  {author} {\bibinfo {author} {\bibfnamefont {H.}~\bibnamefont
  {Bethe}},\ }\href@noop {} {\bibfield  {journal} {\bibinfo  {journal} {Z.
  Phys.}\ }\textbf {\bibinfo {volume} {71}},\ \bibinfo {pages} {931} (\bibinfo
  {year} {1931})}\BibitemShut {NoStop}%
\bibitem [{\citenamefont {Orbach}(1958)}]{orbach_linear_1958}%
  \BibitemOpen
  \bibfield  {author} {\bibinfo {author} {\bibfnamefont {R.}~\bibnamefont
  {Orbach}},\ }\href {\doibase 10.1103/PhysRev.112.309} {\bibfield  {journal}
  {\bibinfo  {journal} {Phys. Rev.}\ }\textbf {\bibinfo {volume} {112}},\
  \bibinfo {pages} {309} (\bibinfo {year} {1958})}\BibitemShut {NoStop}%
\bibitem [{\citenamefont {Korepin}\ \emph {et~al.}(1993)\citenamefont
  {Korepin}, \citenamefont {Bogoliubov},\ and\ \citenamefont
  {Izergin}}]{korepin_quantum_1993}%
  \BibitemOpen
  \bibfield  {author} {\bibinfo {author} {\bibfnamefont {V.~E.}\ \bibnamefont
  {Korepin}}, \bibinfo {author} {\bibfnamefont {N.~M.}\ \bibnamefont
  {Bogoliubov}}, \ and\ \bibinfo {author} {\bibfnamefont {A.~G.}\ \bibnamefont
  {Izergin}},\ }\href
  {http://ebooks.cambridge.org/ebook.jsf?bid=CBO9780511628832} {\emph {\bibinfo
  {title} {Quantum {Inverse} {Scattering} {Method} and {Correlation}
  {Functions}}}}\ (\bibinfo  {publisher} {Cambridge University Press},\
  \bibinfo {address} {Cambridge},\ \bibinfo {year} {1993})\BibitemShut
  {NoStop}%
\bibitem [{\citenamefont {Berry}\ and\ \citenamefont
  {Tabor}(1977)}]{m._v._berry_level_1977}%
  \BibitemOpen
  \bibfield  {author} {\bibinfo {author} {\bibfnamefont {M.~V.}\ \bibnamefont
  {Berry}}\ and\ \bibinfo {author} {\bibfnamefont {M.}~\bibnamefont {Tabor}},\
  }\href {\doibase 10.1098/rspa.1977.0140} {\bibfield  {journal} {\bibinfo
  {journal} {Proc. Roy. Soc. A}\ }\textbf {\bibinfo {volume} {356}},\ \bibinfo
  {pages} {375} (\bibinfo {year} {1977})}\BibitemShut {NoStop}%
\bibitem [{\citenamefont {D'Alessio}\ \emph {et~al.}(2016)\citenamefont
  {D'Alessio}, \citenamefont {Kafri}, \citenamefont {Polkovnikov},\ and\
  \citenamefont {Rigol}}]{dalessio_quantum_2016}%
  \BibitemOpen
  \bibfield  {author} {\bibinfo {author} {\bibfnamefont {L.}~\bibnamefont
  {D'Alessio}}, \bibinfo {author} {\bibfnamefont {Y.}~\bibnamefont {Kafri}},
  \bibinfo {author} {\bibfnamefont {A.}~\bibnamefont {Polkovnikov}}, \ and\
  \bibinfo {author} {\bibfnamefont {M.}~\bibnamefont {Rigol}},\ }\href
  {\doibase 10.1080/00018732.2016.1198134} {\bibfield  {journal} {\bibinfo
  {journal} {Adv. Phys.}\ }\textbf {\bibinfo {volume} {65}},\ \bibinfo {pages}
  {239} (\bibinfo {year} {2016})}\BibitemShut {NoStop}%
\bibitem [{\citenamefont {Atas}\ \emph {et~al.}(2013)\citenamefont {Atas},
  \citenamefont {Bogomolny}, \citenamefont {Giraud},\ and\ \citenamefont
  {Roux}}]{atas_distribution_2013}%
  \BibitemOpen
  \bibfield  {author} {\bibinfo {author} {\bibfnamefont {Y.~Y.}\ \bibnamefont
  {Atas}}, \bibinfo {author} {\bibfnamefont {E.}~\bibnamefont {Bogomolny}},
  \bibinfo {author} {\bibfnamefont {O.}~\bibnamefont {Giraud}}, \ and\ \bibinfo
  {author} {\bibfnamefont {G.}~\bibnamefont {Roux}},\ }\href {\doibase
  10.1103/PhysRevLett.110.084101} {\bibfield  {journal} {\bibinfo  {journal}
  {Phys. Rev. Lett.}\ }\textbf {\bibinfo {volume} {110}},\ \bibinfo {pages}
  {084101} (\bibinfo {year} {2013})}\BibitemShut {NoStop}%
\bibitem [{\citenamefont {Rigol}\ \emph {et~al.}(2008)\citenamefont {Rigol},
  \citenamefont {Dunjko},\ and\ \citenamefont
  {Olshanii}}]{rigol_thermalization_2008}%
  \BibitemOpen
  \bibfield  {author} {\bibinfo {author} {\bibfnamefont {M.}~\bibnamefont
  {Rigol}}, \bibinfo {author} {\bibfnamefont {V.}~\bibnamefont {Dunjko}}, \
  and\ \bibinfo {author} {\bibfnamefont {M.}~\bibnamefont {Olshanii}},\ }\href
  {\doibase 10.1038/nature06838} {\bibfield  {journal} {\bibinfo  {journal}
  {Nature}\ }\textbf {\bibinfo {volume} {452}},\ \bibinfo {pages} {854}
  (\bibinfo {year} {2008})}\BibitemShut {NoStop}%
\bibitem [{\citenamefont {Steinigeweg}\ \emph {et~al.}(2013)\citenamefont
  {Steinigeweg}, \citenamefont {Herbrych},\ and\ \citenamefont
  {Prelov{\v{s}}ek}}]{steinigeweg_eigenstate_2013}%
  \BibitemOpen
  \bibfield  {author} {\bibinfo {author} {\bibfnamefont {R.}~\bibnamefont
  {Steinigeweg}}, \bibinfo {author} {\bibfnamefont {J.}~\bibnamefont
  {Herbrych}}, \ and\ \bibinfo {author} {\bibfnamefont {P.}~\bibnamefont
  {Prelov{\v{s}}ek}},\ }\href {\doibase 10.1103/PhysRevE.87.012118} {\bibfield
  {journal} {\bibinfo  {journal} {Phys. Rev. E}\ }\textbf {\bibinfo {volume}
  {87}},\ \bibinfo {pages} {012118} (\bibinfo {year} {2013})}\BibitemShut
  {NoStop}%
\bibitem [{\citenamefont {Alba}(2015)}]{alba_eigenstate_2015}%
  \BibitemOpen
  \bibfield  {author} {\bibinfo {author} {\bibfnamefont {V.}~\bibnamefont
  {Alba}},\ }\href {\doibase 10.1103/PhysRevB.91.155123} {\bibfield  {journal}
  {\bibinfo  {journal} {Phys. Rev. B}\ }\textbf {\bibinfo {volume} {91}},\
  \bibinfo {pages} {155123} (\bibinfo {year} {2015})}\BibitemShut {NoStop}%
\bibitem [{\citenamefont {Gaudin}(2014)}]{gaudin_bethe_2014}%
  \BibitemOpen
  \bibfield  {author} {\bibinfo {author} {\bibfnamefont {M.}~\bibnamefont
  {Gaudin}},\ }\href
  {http://public.eblib.com/EBLPublic/PublicView.do?ptiID=1582559} {\emph
  {\bibinfo {title} {{T}he {Bethe} {Wavefunction}}}}\ (\bibinfo  {publisher}
  {Cambridge University Press},\ \bibinfo {address} {Cambridge},\ \bibinfo
  {year} {2014})\ \bibinfo {note} {translated by J.-S. Caux}\BibitemShut
  {NoStop}%
\bibitem [{\citenamefont {Essler}\ \emph {et~al.}(2005)\citenamefont {Essler},
  \citenamefont {Frahm}, \citenamefont {G\"ohmann}, \citenamefont {Kl\"umper},\
  and\ \citenamefont {Korepin}}]{essler_one-dimensional_2005}%
  \BibitemOpen
  \bibfield  {author} {\bibinfo {author} {\bibfnamefont {F.~H.}\ \bibnamefont
  {Essler}}, \bibinfo {author} {\bibfnamefont {H.}~\bibnamefont {Frahm}},
  \bibinfo {author} {\bibfnamefont {F.}~\bibnamefont {G\"ohmann}}, \bibinfo
  {author} {\bibfnamefont {A.}~\bibnamefont {Kl\"umper}}, \ and\ \bibinfo
  {author} {\bibfnamefont {V.~E.}\ \bibnamefont {Korepin}},\ }\href
  {http://bilder.buecher.de/zusatz/14/14956/14956838_vorw_1.pdf} {\emph
  {\bibinfo {title} {The one-dimensional {Hubbard} model}}},\ Vol.\ \bibinfo
  {volume} {690}\ (\bibinfo  {publisher} {Cambridge University Press
  Cambridge},\ \bibinfo {year} {2005})\BibitemShut {NoStop}%
\end{thebibliography}%


\begin{thebibliography}{2}%
\makeatletter
\providecommand \@ifxundefined [1]{%
 \@ifx{#1\undefined}
}%
\providecommand \@ifnum [1]{%
 \ifnum #1\expandafter \@firstoftwo
 \else \expandafter \@secondoftwo
 \fi
}%
\providecommand \@ifx [1]{%
 \ifx #1\expandafter \@firstoftwo
 \else \expandafter \@secondoftwo
 \fi
}%
\providecommand \natexlab [1]{#1}%
\providecommand \enquote  [1]{``#1''}%
\providecommand \bibnamefont  [1]{#1}%
\providecommand \bibfnamefont [1]{#1}%
\providecommand \citenamefont [1]{#1}%
\providecommand \href@noop [0]{\@secondoftwo}%
\providecommand \href [0]{\begingroup \@sanitize@url \@href}%
\providecommand \@href[1]{\@@startlink{#1}\@@href}%
\providecommand \@@href[1]{\endgroup#1\@@endlink}%
\providecommand \@sanitize@url [0]{\catcode `\\12\catcode `\$12\catcode
  `\&12\catcode `\#12\catcode `\^12\catcode `\_12\catcode `\%12\relax}%
\providecommand \@@startlink[1]{}%
\providecommand \@@endlink[0]{}%
\providecommand \url  [0]{\begingroup\@sanitize@url \@url }%
\providecommand \@url [1]{\endgroup\@href {#1}{\urlprefix }}%
\providecommand \urlprefix  [0]{URL }%
\providecommand \Eprint [0]{\href }%
\providecommand \doibase [0]{http://dx.doi.org/}%
\providecommand \selectlanguage [0]{\@gobble}%
\providecommand \bibinfo  [0]{\@secondoftwo}%
\providecommand \bibfield  [0]{\@secondoftwo}%
\providecommand \translation [1]{[#1]}%
\providecommand \BibitemOpen [0]{}%
\providecommand \bibitemStop [0]{}%
\providecommand \bibitemNoStop [0]{.\EOS\space}%
\providecommand \EOS [0]{\spacefactor3000\relax}%
\providecommand \BibitemShut  [1]{\csname bibitem#1\endcsname}%
\let\auto@bib@innerbib\@empty
\bibitem [{\citenamefont {Grifoni}\ and\ \citenamefont
  {H{\"{a}}nggi}(1998)}]{grifoni_driven_1998}%
  \BibitemOpen
  \bibfield  {author} {\bibinfo {author} {\bibfnamefont {M.}~\bibnamefont
  {Grifoni}}\ and\ \bibinfo {author} {\bibfnamefont {P.}~\bibnamefont
  {H{\"{a}}nggi}},\ }\href {\doibase 10.1016/S0370-1573(98)00022-2} {\bibfield
  {journal} {\bibinfo  {journal} {Phys. Rep.}\ }\textbf {\bibinfo {volume}
  {304}},\ \bibinfo {pages} {229} (\bibinfo {year} {1998})}\BibitemShut
  {NoStop}%
\bibitem [{\citenamefont {D'Alessio}\ and\ \citenamefont
  {Rigol}(2014)}]{dalessio_long-time_2014}%
  \BibitemOpen
  \bibfield  {author} {\bibinfo {author} {\bibfnamefont {L.}~\bibnamefont
  {D'Alessio}}\ and\ \bibinfo {author} {\bibfnamefont {M.}~\bibnamefont
  {Rigol}},\ }\href {\doibase 10.1103/PhysRevX.4.041048} {\bibfield  {journal}
  {\bibinfo  {journal} {Phys. Rev. X}\ }\textbf {\bibinfo {volume} {4}},\
  \bibinfo {pages} {041048} (\bibinfo {year} {2014})}\BibitemShut {NoStop}%
\end{thebibliography}%

\end{document}


\appendix

\newpage

\title{Supplementary material}
\maketitle
\section{Energies in a Floquet system}
In this appendix, we highlight the different definitions of energy in a Floquet system for a two-step driving protocol
\begin{equation}\label{app:timedepHam}
\hat{H}(t) = 
  \begin{cases}
 \hat{H}_1 &\text{for}\qquad 0 < t < \eta T, \\
  \hat{H}_2    & \text{for}\qquad \eta T < t < T,
  \end{cases}
\end{equation}
with $\hat{H}(t+T) = \hat{H}(t)$. The Floquet operator and Floquet Hamiltonian are subsequently defined as
\begin{equation}\label{app:FloquetHam}
\hat{U}_{F} \equiv  e^{-i \hat{H}_F T} \equiv e^{-i(1-\eta)\hat{H}_2 T} e^{-i \eta \hat{H}_1 T},
\end{equation}
and can be simultaneously diagonalized as
\begin{align}
\hat{H}_F = \sum_n \epsilon_n \ket{\phi_n}\bra{\phi_n}, \ \
\hat{U}_F = \sum_n e^{-i \theta_n} \ket{\phi_n}\bra{\phi_n},
\end{align}
where $\epsilon_n = \theta_n/T$. In Ref. \onlinecite{grifoni_driven_1998}, it was shown how
\begin{equation}\label{app:intHam}
\frac{\partial \theta_n}{\partial T} = \frac{1}{T} \int_{0}^T \braket{\phi_n(t) | \hat{H}(t) | \phi_n(t)} \mathrm{d}t ,
\end{equation}
which is the average energy of a Floquet state during one driving cycle and was shown there to act as a dynamical contribution to the Floquet phase $\theta_n$. For the Floquet operator (\ref{app:FloquetHam}), this can be further simplified, combining
\begin{equation}\label{app:idtUF}
i \partial_T \hat{U}_F = (1-\eta)\hat{H}_2 \hat{U}_F + \eta \hat{U}_F  \hat{H}_1,
\end{equation}
with the Hellmann-Feynman theorem,
\begin{equation}
 \frac{\partial \theta_n}{\partial T} = i \frac{\partial_T \braket{\phi_n | \hat{U}_F |\phi_n}}{\braket{\phi_n | \hat{U}_F |\phi_n}} = \frac{ \bra{\phi_n} i \partial_T \hat{U}_F\ket{\phi_n}}{\bra{\phi_n}\hat{U}_F\ket{\phi_n}}.
\end{equation}
Making use of Eq. (\ref{app:idtUF}) and $\hat{U}_F \ket{\phi_n} = e^{-i \theta_n}\ket{\phi_n}$, this simplifies to
\begin{equation}
\frac{\partial \theta_n}{\partial T} = \bra{\phi_n}(1-\eta)\hat{H}_2+ \eta \hat{H}_1 \ket{\phi_n} = \bra{\phi_n} \hat{H}_{Avg} \ket{\phi_n}.
\end{equation}

Alternatively, this also follows from Eq. (\ref{app:intHam}) by considering the explicit time evolution of the Floquet states $\ket{\phi_n} \equiv \ket{\phi_n(t=0)}$, as governed by 
\begin{equation}
\ket{\phi_n(t)} = \hat{P}(t) \ket{\phi_n} = \hat{U}(t) e^{i \hat{H}_F t} \ket{\phi_n},
\end{equation}
where the evolution operator follows from Eq. (\ref{app:timedepHam}) as
\begin{equation}
\hat{U}(t) = 
  \begin{cases}
 e^{-i \hat{H}_1 t} &\text{for}\ \ 0 < t < \eta T, \\
e^{-i \hat{H}_2 (t-\eta T) } e^{-i \eta \hat{H}_1 T}     & \text{for}\ \  \eta T < t < T.
  \end{cases}
\end{equation}
The two-step driving again allows for a simplification as
\begin{equation}
\hat{U}(t) = 
  \begin{cases}
 e^{-i \hat{H}_1 t} &\text{for}\ \ 0 < t < \eta T, \\
e^{i \hat{H}_2 (T-t) } e^{-i \hat{H}_F T}     & \text{for}\ \  \eta T < t < T.
  \end{cases}
\end{equation}
The kick operator is subsequently given by
\begin{equation}
\hat{P}(t) = 
  \begin{cases}
 e^{-i \hat{H}_1 t} e^{i \hat{H}_F t}&\text{for}\ \ 0 < t < \eta T, \\
e^{i \hat{H}_2 (T-t) } e^{-i \hat{H}_F (T-t)}    & \text{for}\ \  \eta T < t < T,
  \end{cases}
\end{equation}
and the time-evolved eigenstates of the Floquet operator by
\begin{equation}
\ket{\phi_n(t)} = 
  \begin{cases}
 e^{i \epsilon_n t } e^{-i\hat{H}_1t} \ket{\phi_n} &\text{for}\ \ 0 < t < \eta T, \\
e^{-i \epsilon_n (T-t)} e^{i\hat{H}_2(T-t)} \ket{\phi_n}     & \text{for}\ \  \eta T < t < T.
  \end{cases}
\end{equation}
This has a clear interpretation because of the simplicity of the driving protocol. In order to obtain the state in the first part of the period ($0 < t < \eta T$), it is possible to evolve the state forward in time from $t=0$ using only $\hat{H}_1$. For the second half of the period ($\eta T < t < T$), it is possible to evolve the state back in time starting from $t=T$ using only $\hat{H}_2$. This then results in
\begin{equation}
\braket{\phi_n(t)|\hat{H}(t)|\phi_n(t)} = \braket{\phi_n|\hat{H}(t)|\phi_n},
\end{equation}
where inserting this equality in Eq. (\ref{app:intHam}) again returns the time-averaged Hamiltonian.

Given the Floquet phases $\theta_n$, these thus allow for two different measures of the energy of a Floquet state,
\begin{equation}
\frac{\theta_n}{T}=\bra{\phi_n} \hat{H}_F \ket{\phi_n}, \qquad \frac{\partial \theta_n}{\partial T} = \bra{\phi_n}\hat{H}_{Avg} \ket{\phi_n}.
\end{equation}
The derivatives of the phases w.r.t. the period have an interpretation as the average energy of a Floquet state during one driving cycle. These follow from the expectation values of the time-averaged Hamiltonian, and are as such uniquely defined and bounded by the extremal eigenvalues of $\hat{H}_{Avg}$. These can be contrasted to the quasienergies, which are the ratio of the phases and the period, only defined modulo $\frac{2 \pi}{T}$, and commonly taken to be restricted to a single Brillouin (Floquet) zone $\left[-\frac{\pi}{T}, \frac{\pi}{T}\right]$.

\begin{figure}
\includegraphics{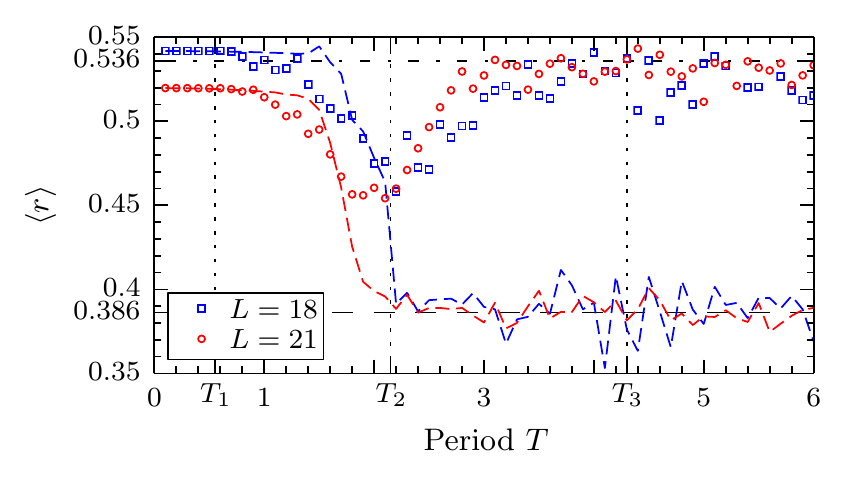}
\caption{Average values $\langle r(\theta_n/T) \rangle$ (symbols) and $\langle r(\partial_T \theta_n) \rangle$ (dashed lines) for periodic driving with $J=1$, $J'=0.8$ and $\Delta_{1,2} = -2.4, -1.6$ and different system sizes $L$.  \label{fig:r_nonint} }
\vspace{-\baselineskip}
\end{figure}

As mentioned in the main text, the Magnus expansion implies that these coincide at small driving periods $T$. For completeness, this is illustrated for a non-integrable $\hat{H}_{Avg}$, following from periodic driving between Hamiltonians
\begin{align}
\hat{H}(t) =& -J \sum_{i} \left[ S_i^x S_{i+1}^x+S_i^y S_{i+1}^y + \Delta(t) S_i^z S_{i+1}^z \right]\nonumber \\
& \qquad + J' \sum_i S_i^z S_{i+2}^z,
\end{align}
where the integrability has been broken by introducing a next-to-nearest-neighbour interaction. For numerical purposes, the calculations are again restricted to periodic boundary conditions and total quasimomentum $k=0$, magnetization $m_z = 1/3$, and parity $p=+1$. These parameters have been chosen in order to make the correspondence with previously obtained results in Ref. \onlinecite{dalessio_long-time_2014}, where three relevant values of the driving period were identified as $T_1=\frac{2 \pi}{W}$, $T_2 = \frac{\pi}{\sigma}$ and $T_3 = \frac{\pi}{\sigma}$. Here, $W$ is the bandwidth of $\hat{H}_{Avg}$ and $\sigma$ is the variance of its spectrum, as elaborated on in the main text. For a given set of ordered levels $\{E_n\}$ the average value of the level spacing ratios 
\begin{equation}
r = \frac{\text{min}(s_n,s_{n+1})}{\text{max}(s_n,s_{n+1})} \in [0,1], \ \  s_n = E_{n+1}-E_{n},
\end{equation}
is given in Fig. \ref{fig:r_nonint} for both quasienergies and averaged energies. While the quasienergies exhibit the behaviour already noted in Ref. \onlinecite{dalessio_long-time_2014}, always remaining close to the GOE statistics of non-integrable Hamiltonians, the average energies exhibit a relatively smooth crossover from GOE to POI statistics between $T_1$ and $T_2$. The POI statistics for the averaged energies are expected when the Floquet Hamiltonian deviates strongly from the time-averaged Hamiltonian. These would then behave as the expectation values of a random local operator ($\hat{H}_{Avg}$) w.r.t. the eigenstates of a (non-)integrable Hamiltonian ($\hat{H}_F$), where subsequent expectation values are not expected to be correlated, resulting in the observed POI statistics.

\section{Perturbation theory}
For two-step periodic driving, the Magnus expansion reduces to the Baker-Campbell-Hausdorff expansion, providing a series expansion of the Floquet Hamiltonian in the driving period $T$ as $\hat{H}_F = \sum_{n=0}^{\infty}\hat{H}_F^{(n)}$. The first three orders simplify significantly when expressed in the time-averaged Hamiltonian and a perturbation term
\begin{equation}
\hat{H}_{Avg} = \eta \hat{H}_1 + (1-\eta)\hat{H}_2, \  \hat{V} = \eta \hat{H}_1 - (1-\eta)\hat{H}_2,
\end{equation}
and can easily be obtained as
\begin{align}
\hat{H}_F^{(0)}&=\hat{H}_{Avg}, \qquad \hat{H}_F^{(1)} = -\frac{iT}{4}[\hat{H}_{Avg},\hat{V}], \nonumber\\
\hat{H}_F^{(2)} &=-\frac{T^2}{24}[[\hat{H}_{Avg},\hat{V}],\hat{V}].
\end{align}
Second-order perturbation theory can now be applied to this Hamiltonian, considering the time-averaged Hamiltonian as the unperturbed Hamiltonian and $T$ as the perturbation strength. The eigenvalues of $\hat{H}_F$ can then be approximated as $\epsilon_n = \epsilon_n^{(0)} + \epsilon_n^{(1)} + \epsilon_n^{(2)} + \mathcal{O}(T^3)$ with
\begin{align}
\epsilon_n^{(1)} &= \bra{\phi_n^{(0)}} \hat{H}_F^{(1)} \ket{\phi_n^{(0)}}, \nonumber \\
\epsilon_n^{(2)} &= \bra{\phi_n^{(0)}} \hat{H}_F^{(2)} \ket{\phi_n^{(0)}} + \sum_{k \neq n} \frac{|\bra{\phi_k^{(0)}} \hat{H}_F^{(1)} \ket{\phi_n^{(0)}}|^2}{\epsilon_n^{(0)}-\epsilon_k^{(0)}},
\end{align}
Because of the commutator structure of the perturbation terms the first-order correction vanishes, $\epsilon_n^{(1)} =0$, whereas the summation in the second-order contribution can be explicitly evaluated as
\begin{align}
&\frac{T^2}{16}\sum_{k \neq n}\frac{|\bra{\phi_k^{(0)}} [\hat{H}_{Avg},\hat{V}] \ket{\phi_n^{(0)}}|^2}{\epsilon_n^{(0)}-\epsilon_k^{(0)}} \nonumber \\
&\qquad =\frac{T^2}{16} \sum_{k \neq n} (\epsilon_n^{(0)}-\epsilon_k^{(0)}) |\bra{\phi_k^{(0)}} \hat{V} \ket{\phi_n^{(0)}}|^2 \nonumber \\ 
& \qquad= \frac{T^2}{16} \sum_k \bra{\phi_n^{(0)}} \hat{V}(\epsilon_n^{(0)}-\hat{H}_{Avg}) \ket{\phi_k^{(0)}}  \bra{\phi_k^{(0)}} \hat{V} \ket{\phi_n^{(0)}} \nonumber \\
& \qquad= \frac{T^2}{16} \bra{\phi_n^{(0)}} \hat{V} (\epsilon_n^{(0)}-\hat{H}_{Avg}) \hat{V} \ket{\phi_n^{(0)}} \nonumber \\
& \qquad= \frac{T^2}{32} \bra{\phi_n^{(0)}} [[\hat{H}_{Avg},\hat{V}],\hat{V}] \ket{\phi_n^{(0)}},
\end{align}
resulting in a total second-order contribution
\begin{align}
\epsilon_n^{(2)} &= -\frac{T^2}{96}\bra{\phi_n^{(0)}}[[\hat{H}_{Avg},\hat{V}],\hat{V}]\ket{\phi_n^{(0)}} \nonumber \\
&= \frac{T^2}{48}\bra{\phi_n^{(0)}}\hat{V} (\hat{H}_{Avg}-\epsilon_n^{(0)}) \hat{V}\ket{\phi_n^{(0)}}.
\end{align}
The dominant correction on the quasienergies is quadratic in the period, and depends not only on the size of the perturbation, but also on its commutator with the time-averaged Hamiltonian. This is a direct result of the Floquet theorem -- if $[\hat{H}_{Avg},\hat{V}]=0$, then $[\hat{H}_1, \hat{H}_2]=0$. The driving then becomes trivial, and the total Floquet Hamiltonian is exactly the time-averaged Hamiltonian for all values of $T$.

\bibliography{FloquetBib.bib}